\documentclass[aps,prb,twocolumn,superscriptaddress,citeautoscript,showpacs,floats]{revtex4}

\usepackage{amsmath}
\usepackage{amssymb}
\usepackage{epsfig}
\usepackage{times}

\graphicspath{{.}{./EPS/}}

\newcommand* {\ket}[1]{\ensuremath{| {#1} \rangle}}
\newcommand* {\ee}{\ensuremath{\mathrm{e}}}

\begin{document}
\title{Effects of a quantum measurement on the electric conductivity:
Application to graphene}

\author{J. Z. Bern\'ad}
\email{j.z.bernad@massey.ac.nz}
\affiliation{Institute of Fundamental Sciences and MacDiarmid Institute for Advanced
Materials and Nanotechnology, Massey University (Manawatu Campus), Private Bag
11~222, Palmerston North 4442, New Zealand}

\author{M. J\"a\"askel\"ainen}
\affiliation{Institute of Fundamental Sciences and MacDiarmid Institute for Advanced
Materials and Nanotechnology, Massey University (Manawatu Campus), Private Bag
11~222, Palmerston North 4442, New Zealand}

\author{U. Z\"ulicke}
\affiliation{Institute of Fundamental Sciences and MacDiarmid Institute for Advanced
Materials and Nanotechnology, Massey University (Manawatu Campus), Private Bag
11~222, Palmerston North 4442, New Zealand}
\affiliation{Centre for Theoretical Chemistry and Physics, Massey University (Albany
Campus), Private Bag 102904, North Shore MSC, Auckland 0745, New Zealand}
\date{\today}

\begin{abstract}
We generalize the standard linear-response (Kubo) theory to obtain the conductivity
of a system that is subject to a quantum measurement of the current. Our
approach can be used to specifically elucidate how back-action inherent to quantum
measurements affects electronic transport. To illustrate the utility of our general formalism, we calculate the
frequency-dependent conductivity of graphene and discuss the effect of
measurement-induced decoherence on its value in the dc limit. We are able to
resolve an ambiguity related to the parametric dependence of the minimal
conductivity.
\end{abstract}


\maketitle

The fact that measurements exert a back-action on the measured object has 
attracted a lot of attention~\cite{Braginski1992,Diosi2006,JacobsSteck2007}, partly
due to its relevance for the foundations of quantum physics, but also
because of implications for metrology~\cite{Geremia2003} and the design of
solid-state devices~\cite{Bernad2008}. Fundamental considerations necessitate a 
distinction between selective and nonselective descriptions of quantum 
measurements~\cite{Mensky1998}.\ {\em Selective\/} descriptions use stochastic
differential equations~\cite{Gisin1984,CavesMilburn1987}, or restricted path
integrals~\cite{Mensky1979}, and result in conditional quantum dynamics when the
measurement results are recorded. Some properties of selective measurements
have been verified experimentally for areas as diverse as cavity
QED~\cite{Guerlin2007} and superconducting phase qubits~\cite{Katz2006}.\
{\em Nonselective\/} descriptions represent the evolution of the measured system
irrespective of the measurement result. This description takes into account all
possible readouts, and the actual readout is assumed not to be
known~\cite{CavesMilburn1987,Mensky1998}. Quantum-mechanical back-action on 
the unsharply measured system causes loss of coherence between eigenstates 
of the measured quantity.
In this work, we discuss measurement
back-action theoretically within the nonselective framework. This approach makes it possible to determine how a macroscopic
observable such as the conductivity of a system is affected when the current is
detected in an unsharp-measurement scenario. We derive a generalized Kubo
formula where the measurement back-action provides a natural damping
mechanism.
We demonstrate the power of the developed general formalism by calculating the
frequency-dependent conductivity $\sigma(\omega)$ of graphene~\cite{Novoselov},
a promising candidate for future micro- and nanoelectronics ~\cite{Scalability} and
also a low energy laboratory of relativistic physics ~\cite{YZhang,Novoselov2005}.
Electronic-transport properties of graphene were analyzed in several previous
studies using different methods; e.g., the Landauer-B\"uttiker
formalism~\cite{Landauer,Ryu}, the linear-response Kubo
formula~\cite{KuboWorks,Ziegler1,Ziegler2,Ryu}, and the Boltzmann
equation~\cite{Boltzmann}. It was found that different parametric dependences of
the dc conductivity can result from different limiting procedures applied to ordinary
Kubo formulae~\cite{Ziegler2}. Within our generalized Kubo formalism, we obtain
physical conditions for when these results apply. Two regimes can be distinguished 
by comparison of the energy scale $\hbar\Gamma$ that quantifies
measurement-induced decoherence with the greater one among the thermal energy
$k_{\text{B}} T$ and the chemical potential $\mu$ (measured from the Dirac point).
Weak back-action ($\hbar\Gamma<\text{max}\{k_{\text{B}} T, \mu\}$) results in
Drude-type behavior $\sigma(0)\propto 1/\Gamma$. In the opposite
(strong-back-action) limit, a mixing of intra-band and inter-band contributions occurs
that changes the parametric dependence of the dc conductivity such that
$\sigma(0)\propto\Gamma$.

We employ the linear-response (Kubo) formalism~\cite{Madelung} and
divide the system's Hamiltonian into the part $\hat{H_0}$, which governs the free
evolution, and $\delta\hat{H}$, the perturbation associated with an external electric
field $\mathbf{E}$. For simplicity, we take the latter to be constant in space and
assume the field to be applied between $t=-\infty$ and $t=0$. The perturbation
Hamiltonian is $\delta \hat{H}=-e \mathbf{E}\cdot \hat{\mathbf{r}}\,\ee^{i \omega t}$.
In a nonselective description~\cite{Diosi2006,JacobsSteck2007}, the dynamics of
the density matrix, when the current is measured, is governed by a master equation with the back-action caused by 
a term of Lindblad form~\cite{POVM} 
\begin{equation}\label{mastermeas}
\frac{d\hat{\rho}}{dt}=-\frac{i}{\hbar}[\hat{H},\hat{\rho}]-\frac{\gamma}{8}[\hat{j},[\hat{j},\hat{\rho}]]=\mathcal{L}\hat{\rho}-\frac{i}{\hbar}[\delta\hat{H},\hat{\rho}] ,
\end{equation}
where $\mathcal{L}\hat{\rho}=-\frac{i}{\hbar}[\hat{H}_0,\hat{\rho}]-\frac{\gamma}{8}[\hat{j},[\hat{j},\hat{\rho}]]$ with the current operator
$\hat{j}=i\frac{e}{\hbar}[\hat{H}_0,\hat{r}]$ being our measured observable\cite{OperatorNote}. We note that the measured observable is arbitrarily chosen to be the current, as it is relevant for the conductance calculation at hand.
The main parameter of an unsharp 
quantum measurement device, as described by  Eq.~(\ref{mastermeas}), is the 
detection performance $\gamma=(\Delta t)^{-1}(\Delta j)^{-2}$, where $\Delta t$ is 
the time resolution or, equivalently, the inverse bandwidth of the detector, and
$\Delta j$ the statistical error characterizing unsharp detection of the average
current. Within linear-response theory, we can linearize $\hat{\rho}=\hat{\rho}_0+
\delta{\hat{\rho}}$, where $\hat{\rho}_0$ is the system's equilibrium density matrix. Keeping only linear terms in Eq.~(\ref{mastermeas}), we get
\begin{equation}
\frac{d \delta{\hat{\rho}}}{dt}=\mathcal{L}\delta{\rho}-\frac{i}{\hbar}[\delta \hat{H},
\hat{\rho}_0]-\frac{\gamma}{8}[\hat{j},[\hat{j},\hat{\rho}_0]] \, ,
\end{equation} 
assuming that the unsharp detection does not affect the equilibrium and using
$[\hat{H_0},\hat{\rho}_0]=0$ as well as $[\delta \hat{H},\delta \hat{\rho}]\simeq 0$.
Introducing $\Delta \hat{\rho}=\ee^{-\mathcal{L}t}\delta \hat{\rho}$ yields
\begin{equation}
 \frac{d \Delta \hat{\rho}}{dt}=\ee^{-\mathcal{L}t}\left(-\frac{i}{\hbar}[\delta 
\hat{H},\hat{\rho}_0]-\frac{\gamma}{8}[\hat{j},[\hat{j},\hat{\rho}_0]]\right) \, .
\end{equation}
Note that $\Delta \hat{\rho}$ and $\delta \hat{\rho}$ have the same value at $t=0$,
and both vanish at $t=-\infty$. Integration yields
\begin{equation}\label{innerres}
\delta \hat{\rho}(t=0)=\int^{0}_{-\infty}dt \,\,\ee^{-\mathcal{L}t}\left(-\frac{i}{\hbar}
[\delta \hat{H},\hat{\rho}_0]-\frac{\gamma}{8}[\hat{j},[\hat{j},\hat{\rho}_0]]\right) .
\end{equation}
The exponential factor in Eq.~(\ref{innerres}) ensures convergence of the time 
integral, making it unnecessary to introduce the phenomenological adiabatic
damping parameter employed in conventional linear-response
theory~\cite{Madelung}. Inserting (\ref{innerres}) into the expectation value for the
current density $\hat{j}$ and dividing by $|\mathbf{E}|$ yields the optical conductivity 
\begin{equation} \label{mastermeas2}
 \sigma_{\mu \nu}(\omega)=\int_{-\infty}^{0} \left[ K_{\mu\nu}(t) e^{i \omega t} +
 K'_{\mu\nu}(t) \right]\, dt \, , 
\end{equation}
with the kernels
\begin{equation}\label{KKp}
K_{\mu\nu}(t)=-\frac{1}{i\hbar} {\mathrm{Tr}}\left\{ \hat{j}_{\mu} \ee^{-\mathcal{L}
(\hat{j}_{\mu})t}\left( [e\hat{r}_{\nu},\hat{\rho}_{0}]\right)\right\} \, ,
\end{equation}
\begin{equation}\label{KKp2}
K'_{\mu\nu}(t)= {\mathrm{Tr}} \left \{\frac{\hat{j}_{\mu}}{E_{\nu}} \ee^{-\mathcal{L}
(\hat{j}_{\mu})t}\left(-\frac{\gamma}{8}[\hat{j}_{\mu},[\hat{j}_{\mu},\hat{\rho}_0]]\right)
\right\}\, .
\end{equation}

Equations~(\ref{KKp}) and (\ref{KKp2}) are the general result for the response to an
unsharp measurement and could, in principle, be applied to any quantum system. 
As an instructive application, we now use Eq.~(\ref{mastermeas2}) to calculate the
frequency-dependent conductivity of single-layer graphene within the continuum
model for quasiparticles close to the $\mathbf{K}$ point in the Brillouin zone. In
Eq.~(\ref{mastermeas}), the Hamiltonian $\hat{H}_0$ could, in principle, include 
Coulomb-interactions, impurity scattering, density dependencies etc. In the present
work, we just use the free-quasiparticle Hamiltonian for graphene in plane-wave
representation,
\begin{equation}\label{Hamilton}
\hat{H}_0(\mathbf{k})=\hbar v (\sigma_x k_x+\sigma_y k_y) \, ,
\end{equation}
where $k_x, k_y$ are the Cartesian components of wave vector $\mathbf{k}$,
$\sigma_i$ denote Pauli matrices acting in the sublattice-related pseudo-spin space,
and $v$ is the Fermi velocity, which has a value $\simeq 10^6$~m/s.  With the
position operator $\hat{\mathbf r}$ being the wave-vector gradient, the current
operator is $\hat{j}_{\mu}=\frac{ie}{\hbar}[\hat{H}_0(\mathbf{k}),\hat{r}_{\mu}]=
\frac{e}{\hbar} \frac{\partial\hat{H}_0( \mathbf{k})}{\partial k_{\mu}}$.
Single-particle eigenstates of clean graphene can be written as a direct product of a
plane wave in configuration space with a spinor $\ket{n}=\ket{\mathbf{k}} \otimes
\ket{\sigma}_{\mathbf{k}}$. Here $\sigma$ labels the electron and hole bands,
respectively, and the spinor wave function depends on wave vector $\mathbf{k}$. 
The current operators with the Hamiltonian (\ref{Hamilton}) in the spinor space are
$\hat{j}_x= e v \sigma_x$, and $\hat{j}_y= e v \sigma_y$ \cite{superselection}. From 
the definition of the equilibrium density matrix in the spinor space we find
$\hat{\rho}_0 \ket{\sigma}_{\mathbf{k}}=f(\hbar \epsilon_{\mathbf{k},\sigma})
\ket{\sigma}_{\mathbf{k}}$, and $\hat{H}_0 \ket{\sigma}_{\mathbf{k}}=\hbar 
\epsilon_{\mathbf{k},\sigma}\ket{\sigma}_{\mathbf{k}}$, where $f$ is the Fermi-Dirac 
distribution function and $\epsilon_{\mathbf{k},\pm}=\pm |\mathbf{k}|$. (For
simplicity, the speed $v$ has been absorbed into $\mathbf{k}$.) Together with the
anti-symmetry in momentum space, this implies  $K'_{\mu \nu}=0$. The calculation
of $K_{\mu\nu}$ is straightforward, and using the Laplace transform to solve for the
dynamics, we find
\begin{equation}\label{K}
K_{\mu\nu}(t)=\frac{e^2}{\hbar}\int \frac{d^2 \mathbf{k}}{(2 \pi)^2} \mathbf{Res}
\{\tilde{K}_{\mu\nu}(\mathbf{k},\gamma,z)e^{zt}\},
\end{equation}
where $\mathbf{Res}$ stands for the sum of residues of the integrand. 
It can be seen
that $K_{xy}=K_{yx}=0$. The remaining two conductivities are identical, as a change
of variables $k_x \leftrightarrow k_y$ in the expression for $K_{xx}$ yields $K_{yy}$. 
The choice of conductivity measured will decide what kind of back-action will influence the system.
Here we take $\hat{j}_x$, the current along the x-direction, which breaks the isotropy of the problem and we find
\begin{widetext}
\begin{equation}\label{Kfull}
\tilde{K}_{xx}(\mathbf{k},\Gamma,z)=\frac{ k_x^2 \mid \mathbf{k} \mid (16
\Gamma^2-8 z \Gamma+4 \mid \mathbf{k} \mid^2 +z^2 )\sum_{\sigma}\left(-
\frac{df(\hbar \epsilon)}{d\epsilon}\mid_{\epsilon=\epsilon_{\mathbf{k},\sigma}} \right)
+k_y^2 (z-4 \Gamma)^2 \left[ f(\hbar\epsilon_{\mathbf{k},-})-f(\hbar\epsilon_{\mathbf{k},+}) \right]}
{\mid \mathbf{k} \mid^3 \left[ 16 z \Gamma^2-8 (2k_y^2+z^2) \Gamma +z (4 \mid 
\mathbf{k} \mid^2 + z^2) \right] }, 
\end{equation}
\end{widetext}
where the parameter
$\Gamma=\gamma e^2 v^2/8$ was introduced. The cubic factor in the denominator 
of Eq.~(\ref{Kfull}) has three roots $z_i$, which give the poles in Eq.~(\ref{K}).

For small $\Gamma$, the roots are to lowest order $z_1=0$ and $z_{2,3}= \pm 2 i
|\mathbf{k}|$. Using this and performing the time-integration in the limit $\Gamma\to
0$, the known $intra$- and $inter$-band contributions
\begin{eqnarray}\label{clean2}
\frac{\sigma^{(intra)}}{\sigma_0}&=&\frac{\pi}{2}\delta(\omega)
\int_0^{\infty}x\left[\sum_{\sigma=\pm}\left(-\frac{df(\hbar \epsilon)}{d\epsilon}\mid_{\epsilon=\sigma x} \right)\right]dx \nonumber \\
 &=&\frac{\pi}{2}\delta \left( \frac{\hbar \omega}{k_B T} \right) \left[ 2 \log
 \left(1+\ee^{\frac{\mu}{k_B T}}\right)-\frac{\mu}{k_B T}\right] ,
\end{eqnarray}
\begin{equation}\label{clean}
\frac{\sigma^{(inter)}}{\sigma_0}=\frac{\pi}{8}\frac{\sinh(\frac{\hbar \omega}
{2k_BT})}{\cosh(\frac{\mu}{k_BT})+\cosh(\frac{\hbar \omega}{2k_BT})},
\end{equation}
to the conductivity of clean graphene are found. The scale factor
$\sigma_0=4e^2/h$ accounts for spin and pseudospin degeneracy.

\begin{figure}[t]
\begin{center}
\includegraphics[width=8.6cm]{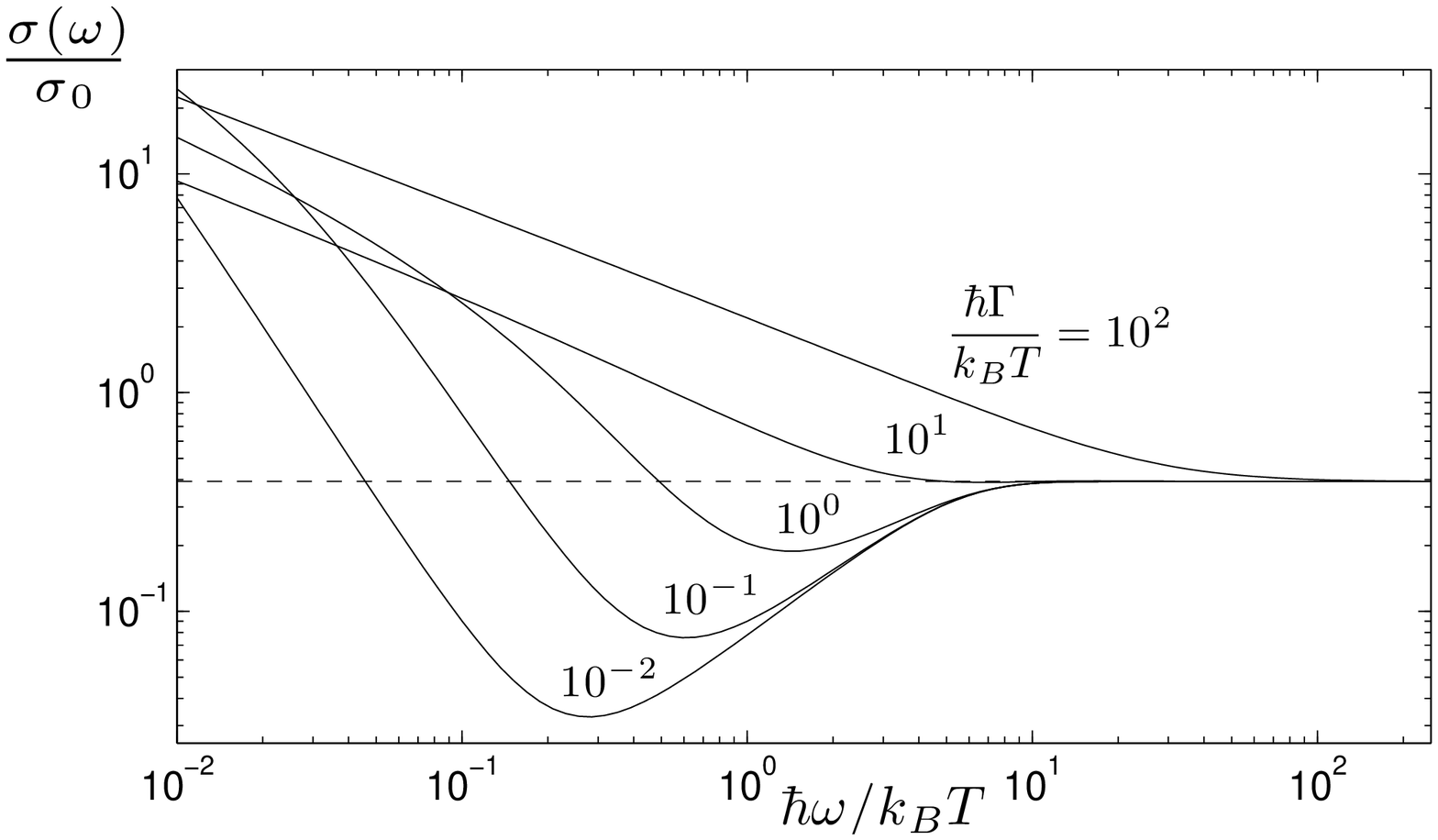}
\caption{\label{omega}Frequency dependence of the conductivity of clean
single-layer graphene when the current is unsharply quantum measured. The 
chemical potential is fixed at $\mu/ k_B T =1$, and the value for
measurement-induced decoherence assumed for each curve is indicated. The
dashed line is at $\pi/8$. As the coupling to the measurement device is decreased,
the curves more closely resemble the result found for clean graphene.}
\end{center}
\end{figure}

\begin{figure}[t]
\begin{center}
\includegraphics[width=8.6cm]{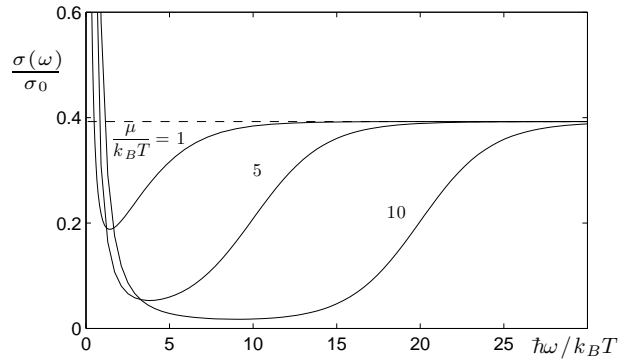}
\caption{\label{mu} AC conductivity of single-layer graphene, for fixed $\hbar
\Gamma/k_B T=1$ and different values of chemical potential $\mu$ (measured
from the Dirac point).}
\end{center}
\end{figure}

\begin{figure}[b]
\begin{center}
\includegraphics[width=8.6cm]{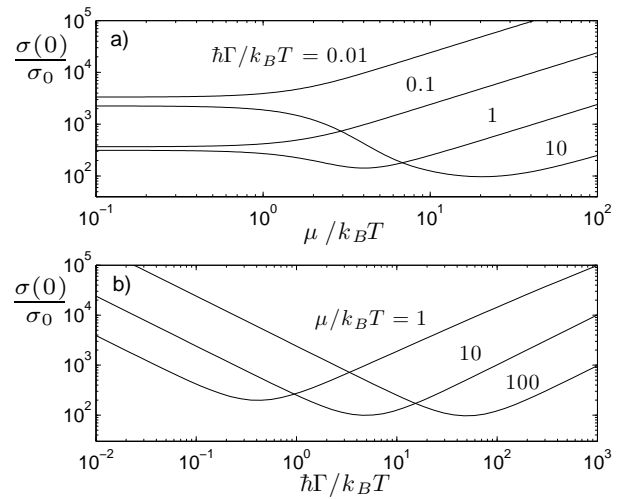}
\caption{\label{zero}
a)~DC conductivity plotted as a function of chemical potential for different 
values of $\Gamma$. The conductivity is linear in the chemical potential for large 
values of $\mu/k_BT$. b)~DC conductivity shown as a function of $\hbar \Gamma
/k_B T$ for different values of the chemical potential.  For $\mu > k_{\text{B}} T$
and $\hbar\Gamma  < \mu$, the conductivity is inversely proportional to the
effective rate of decoherence, in analogy with the behavior expected from 
inelastic impurity scattering. In contrast, for $\hbar \Gamma  > \mu$ there is direct 
proportionality. This behavior was exhibited as long as $\mu\gtrsim k_{\text{B}} T$, 
whereas curves for $\mu\le k_{\text{B}} T$ practically coincide.}
\end{center}
\end{figure}

In the following, the conductivity is calculated numerically for finite values of
$\Gamma$ from Eq.~(\ref{mastermeas2}) with Eqs.~(\ref{K}) and (\ref{Kfull}). 
We use $k_B T$ as unit of energy. Figures~\ref{omega} and \ref{mu} show
the ac (optical) conductivity~\cite{Mak2008}, whereas Fig.~\ref{zero} shows
the dc conductivity that is measured, e.g, in mesoscopic transport experiments.

In Fig.~\ref{omega}, the conductivity is shown as function of frequency for different
values of the coupling strength when the chemical potential remains fixed. For high
frequencies, the conductivity saturates to  the universal value $\pi/8$, indicated by a
dashed line. The detailed shape of the crossover to saturation depends on
$\Gamma$, with higher values pushing it to higher frequencies. 
In the limit of small $\Gamma$, measurement-induced decoherence appears to
simulate the effect of life-time broadening due to inelastic
scattering~\cite{Ziegler1,Ziegler2,Peres2008}, but a closer look reveals that it is 
fundamentally different. Technically, the effect of $\Gamma$ goes beyond merely
broadening of distribution functions, it also moves the position of their peaks in 
energy, thus changing the resonance condition. For clean graphene, there is only 
a delta function peak for the $intra$-band contribution, whereas here the $intra$- 
and $inter$-band contributions to the conductivity become mixed. The existence of 
such a mixing has been inferred from recent experiments~\cite{Mak2008}.
The dependence on the chemical potential is illustrated in Fig.~\ref{mu}. Generally,  
increasing the chemical potential shifts the frequency beyond which the conductivity
attains its universal saturation to higher values. This behavior is as expected
theoretically~\cite{Peres2008} and observed in experiments~\cite{Mak2008}.  At frequencies $\omega$ smaller than a few times the chemical potential, there is a significant departure from the universal conductance plateau. For fixed $\mu$ and $k_{\text{B}}T$ the saturation point in scattering models is independent of the scattering parameter, whereas in our work it strongly depends on the value of $\Gamma$. For $\mu \gtrsim \hbar\Gamma$, the saturation occurs as in clean graphene, whereas the opposite case gives saturation for larger frequencies with increasing $\Gamma$, as seen in Fig.~\ref{omega}.

When the effect of disorder is modeled conventionally by a life-time broadening due
to inelastic scattering~\cite{Ziegler1,Ziegler2}, a Drude peak is found for $\hbar 
\omega/ k_{\text{B}} T \sim 0$, with a height inversely proportional to the
inelastic-scattering rate. Since the master equation describing the effect of quantum
measurements is formally identical to certain models of decoherence, we expect that
$\sigma(0)/\sigma_0 \sim 1/\Gamma$. This turns out to be correct only for $\hbar
\Gamma < \text{max}\{\mu, k_{\text{B}} T\}$,  as can be seen in Fig.~\ref{zero}b. In 
the opposite limit, the parametric dependence of the dc conductivity is changed; it
then grows linearly with $\Gamma$. A similar anomalous behavior was obtained in
Ref.~~\onlinecite{Ziegler2} by applying an unconventional limiting procedure within
the Kubo formalism. Here we are able to readily identify the regimes where
ordinary-Drude behavior or anomalous band-mixing behavior will be exhibited.
Clearly, observing the latter should be easiest when the chemical potential is at the
Dirac point and the temperature is low enough to satisfy $\hbar\Gamma > 
k_{\text{B}}T$, which corresponds to the minimal conductivity regime for graphene. Figure~3a shows how the nonmonotonic $\Gamma$-dependence of
the conductivity would be manifested in a typical transport experiment where the
chemical potential (i.e., the density) is varied.

The theory presented here is based on an unsharp measurement of the current 
density. To estimate the magnitude $\Gamma$ of measurement-induced
decoherence, we must consider the two situations most closely related to our result, 
optical conductivity measurements, and mesoscopic transport measurements. For 
the latter case, it can be easily seen that $e^2 v^2/(\Delta j)^2 $is  the
signal-to-noise ratio. Using typical values $\Delta\nu \gtrsim 10^6 $~Hz for the 
bandwidth and $\Delta I/I \approx 10^{-3}$ for the signal-to-noise ratio, we find that 
values up to $\hbar\Gamma/k_{\text{B}}T \approx 10^{-1}-10^{2}$ can be achieved 
for $T \approx 1 - 300$~K. For measurements of the optical conductivity, the 
situation is more complex, as the measured signal is induced by the currents 
generated by the applied optical field. As a result, additional uncertainties such as geometrical factors and detector efficiencies become important, possibly bringing down the detection performance, but this can in principle be compensated by the increase in bandwidth offered by optical detectors, $\Delta\nu \gtrsim 10^9 $~Hz.

In conclusion, we derived a new Kubo formula to study the effect of
measurement-induced back-action on the conductivity. The back-action naturally 
introduces a source of damping and thus makes the converged adiabaticity 
parameter frequently used in Kubo formula calculations superfluous. We applied this 
approach to calculate the electric conductivity of single-layer graphene. Mixing of the
$intra$- and $inter$-band contributions to the dc conductivity strongly affect its
parametric dependence on the detector performance $\Gamma$. The regime of 
weak coupling to the measuring device models a standard Drude-type behavior, 
whereas in the opposite limit of strong back-action, we find that measuring a current 
in graphene will actually \emph{enhance} the conductivity.

This work was supported by the Massey University Research Fund and by the 
Marsden Fund Council (contract MAU0702) from Government funding, administered
by the Royal Society of New Zealand.


\begin{thebibliography}{99}
\bibitem{POVM} Starting from POVM and using the standard Gaussian model we arrive in the selective measurement case to stochastic master equation and in the non-selective case at a master equation with the double commutator.
\bibitem{OperatorNote} For other types of experiments, the corresponding measured operator will occur in the Lindbladian term and, thus, induce back-action on the system.  Although the formalism is, in its mathematical structure, identical to decoherence, the effects on the dynamics depend on what kind of measurement is performed, which is a matter of choice. In contrast, environmentally induced decoherence that is determined by interaction with an ever-present reservoir
cannot be manipulated at will. The choice of measured observable thus decides the type of  decoherence present in the system through experimental design.
\bibitem{superselection}
The Hilbert space is decomposed into irreducible components. Each pseudo-spin sector is a superselection sector, preserved by the current operator.
\end{thebibliography}
\end{document}